\documentclass{ws-ijmpa}

\begin{document}

\renewcommand{\draftnote}{} 
\renewcommand{\trimmarks}{} 

\markboth{E. A. Matute} {Electroweak Quark--Lepton Symmetry}

\catchline{}{}{}{}{}

\title{ELECTROWEAK QUARK--LEPTON SYMMETRY AND\\WEAK
TOPOLOGICAL-CHARGE CONFINEMENT IN THE STANDARD MODEL WITH DIRAC
NEUTRINOS}

\author{\footnotesize ERNESTO A. MATUTE}

\address{Departamento de F\'{\i}sica, Universidad de Santiago de
Chile, USACH,\\ Casilla 307 -- Correo 2, Santiago, Chile\\
ematute@lauca.usach.cl}

\maketitle

\pub{}{}

\begin{abstract}
The standard electroweak model with Dirac neutrinos is extended by
way of the principles of electroweak quark--lepton symmetry and
weak topological-charge confinement to account for quark--lepton
charge relations which, if not accidental, are indicative of
charge structures. A mixing in quarks and leptons of underlying
integer local charges with integer weak topological charges
associated with an additive group $Z_3$, fixed by the anomaly
cancellation requirement, is discussed. It is found that the
electroweak difference between topological quarks and leptons is
the nonequivalence between the topological vacua of their weak
field configurations, produced by a four-instanton which carries
the topological charge, induces the universal fractional piece of
charge distinguishing quarks from leptons, and breaks the
underlying symmetry. The constituent quarks of the standard model
appear as coming from topological quarks, via the weak
four-instanton event. Dual transitions occur for leptons. It is
shown that several other fundamental problems left open in the
standard electroweak model with Dirac neutrinos are solved: the
one-to-one correspondence between quark and lepton flavors, the
existence of three generations, the conservation and ungauging of
$B-L$, the electric charge quantization, and the confinement of
fractional electric charges.

\end{abstract}

\ccode{PACS numbers: 11.15.Kc, 11.30.Hv, 11.30.Fs, 12.60.-i}

\section{Introduction}
\label{introduction}

In spite of the success of the $\mbox{SU(2)}_{L} \times
\mbox{U(1)}_{Y}$ Standard Model (SM) of electroweak
interactions,\cite{Glashow}\cdash\cite{Salam} there are many
puzzling questions that demand its extension to new physics. We
name the following ones which are somehow related to each other:
the discovery of neutrino mass, the evidence of a one-to-one
correspondence between quark and lepton flavors, the existence of
three families of quarks and leptons, the observed conservation
and ungauging of the baryon minus lepton number, the fact that the
electric charge is quantized, and the confinement of fractional
electric charges. All these questions have been addressed by
ambitious programs, including the theories of grand
unification\cite{Pati,Georgi} and strings,\cite{Scherk} which
invoke physics near the Planck scale.

The purpose of this paper is to answer the above deep questions at
the (low-energy) level of the SM extended with Dirac right-handed
neutrinos.  Here we mention that even though there are evidences
for nonzero neutrino masses from neutrino oscillation
experiments,\cite{Data} no light on their Dirac or Majorana
character has been shed yet; search for Majorana neutrino effects
in double-beta decays, however, currently has no positive results.

As recently noted,\cite{Matute1,Matute2} once Dirac right-handed
neutrinos are included for considering massive neutrinos, an
electroweak quark--lepton symmetry is established, manifested in
part through the following patterns of normalized weak
hypercharge:
\begin{eqnarray}
\begin{array}{rcll}
Y(u_{aL}) &=& \displaystyle Y(\nu_{aL}) + \frac{4}{3} =
\frac{1}{3} \, , \qquad & \displaystyle Y(d_{aL}) = Y(e_{aL})
+ \frac{4}{3} = \frac{1}{3} \, , \\[12pt]
Y(u_{aR}) &=& \displaystyle Y(\nu_{aR}) + \frac{4}{3} =
\frac{4}{3} \, , \qquad & \displaystyle Y(d_{aR}) = Y(e_{aR}) +
\frac{4}{3} = - \frac{2}{3} \, ,
\end{array}
\label{hyper}
\end{eqnarray}
as well as normalized electric charge:
\begin{equation}
Q(u_{a}) = Q(\nu_{a}) + \frac{2}{3} = \frac{2}{3} \, , \qquad
Q(d_{a}) = Q(e_{a}) + \frac{2}{3} = - \frac{1}{3} \, ,
\label{Qcharge}
\end{equation}
where the subscript $a$ denotes the three generations and the
conventional relation
\begin{equation}
Q = T_{3} + \frac{1}{2} \; Y
\label{charge}
\end{equation}
between electric charge and hypercharge is being used.

To capture the true meaning of the global fractional piece of
hypercharge, however, we now claim that this part, which is not
affected by the local electroweak interactions, is just $4/3$
times $-(3B-L)$ and obviously a good global symmetry of the SM,
where $B$ and $L$ are the usual baryon and lepton numbers and the
$3$ is due to the number of quark colors. Thus, the hypercharge of
a quark and its lepton partner are actually related according to
\begin{eqnarray}
\begin{array}{rcl}
Y(q) &=& \displaystyle Y(\ell) - \frac{4}{3} \, (3B-L)(\ell) \, ,
\\[12pt] Y(\ell) &=& \displaystyle Y(q) - \frac{4}{3}
\, (3B-L)(q) \, ,
\end{array}
\label{newhyper}
\end{eqnarray}
where we have suppressed the subscripts $L$, $R$, and used the
property
\begin{equation}
(3B-L)(\ell) = - (3B-L)(q) \, .
\end{equation}
In turn, using Eq.~(\ref{charge}) with the discrete symmetry
$T_{3}(q)=T_{3}(\ell)$ between quark and lepton partners,
Eq.~(\ref{Qcharge}) becomes shaped as
\begin{eqnarray}
\begin{array}{rcl}
Q(q) &=& \displaystyle Q(\ell) - \frac{2}{3} \, (3B-L)(\ell) \, ,
\\[12pt] Q(\ell) &=& \displaystyle Q(q) - \frac{2}{3}
\, (3B-L)(q) \, .
\end{array}
\label{newQcharge}
\end{eqnarray}

On the other hand, the term $4/3$ in Eq.~(\ref{hyper}) can also be
expressed as the difference of $B-L$ between quarks and leptons.
This implies a relation for the charges of quark and lepton
partners having the form
\begin{equation}
Y(q) - (B-L)(q) = Y(\ell) - (B-L)(\ell) \, .
\label{another}
\end{equation}
The corresponding formula in terms of electric charge is
\begin{equation}
Q(q) - \frac{1}{2} \, (B-L)(q) = Q(\ell) - \frac{1}{2} \,
(B-L)(\ell) \, .
\end{equation}

The crucial question that next follows is whether the charge
relations between quark and lepton partners in
Eqs.~(\ref{newhyper}) and (\ref{another}) are real or accidental.
Real in the sense that simple and direct principles on discrete
symmetries exist which lead to these charge connections, somehow
extending to the electromagnetic sector the discrete symmetry
found in the weak sector and known as quark--lepton
symmetry.\cite{Gamba,GIM} Accidental in the sense that the charge
relations are just interplays of independent contributions of the
hypercharge, baryon and lepton quantum numbers assigned to quarks
and leptons, with the rather magical result that the new charge
$Y-4(3B-L)/3$ of a quark (lepton) is identical to the $Y$ of its
lepton (quark) partner, and the one defined as $Y-(B-L)$ is the
same for both of them.

It is hard to believe that the interplay of the $Y$, $B$, and $L$
quantum numbers of quarks and leptons in Eqs.~(\ref{newhyper}) and
(\ref{another}) be accidental.  If the charge relations are real
and simple and direct discrete symmetries are behind, one
concludes on a symmetric charge structure such that each quark
(lepton) has an underlying integer (fractional) charge equal to
that of its lepton (quark) partner.  We refer to these new bare
charge states as prequarks (preleptons) and denote them as
$\hat{q}$ ($\hat{\ell}$). Specifically, from Eq.~(\ref{newhyper})
one has a charge dequantization according to the following form:
\begin{eqnarray}
\begin{array}{rcl}
Y(q) &=& \displaystyle Y(\hat{q}) - \frac{4}{3} \, (B-3L)(\hat{q})
\, , \\[12pt] Y(\ell) &=& \displaystyle Y(\hat{\ell}) -
\frac{4}{3} \, (B-3L)(\hat{\ell}) \, ,
\end{array}
\label{hathyper}
\end{eqnarray}
with prequark--lepton and prelepton--quark charge symmetries given
by
\begin{eqnarray}
\begin{array}{rcl}
Y(\hat{q}) &=& Y(\ell) \, , \hspace{0.7cm} Y(\hat{\ell}) = Y(q) \,
, \\[12pt] (B-3L)(\hat{q}) &=&
(3B-L)(\ell) = -(3B-L)(q) \, , \\[12pt]
(B-3L)(\hat{\ell}) &=& (3B-L)(q) = -(3B-L)(\ell) \, .
\end{array}
\label{hatql}
\end{eqnarray}
In Eq.~(\ref{hathyper}) we have the $4/3$ associated with the
combination $B-3L$ in place of $3B-L$ because prequarks
(preleptons) are the ones that now possess integer (fractional)
charges. The charge $-4(B-3L)/3$ is therefore the global
charge-shift which normalizes the prefermion charges to the SM
values. From Eq.~(\ref{hatql}) we readily obtain $B(\hat{q})=-1$
and $L(\hat{\ell})=-1/3$.  Thus, $B-L$ and not $B-3L$ is the same
for quarks (prequarks) and preleptons (leptons).  We see that the
$B-3L$ in Eq.~(\ref{hathyper}) is essentially a bookkeeping global
charge based on counting such that three preleptons make a system
with an integer hypercharge and one unit of $B-L$ charge, just as
three quarks do.

From Eq. (\ref{another}) instead one has
\begin{eqnarray}
\begin{array}{rcl}
Y(q) - (B-L)(q) &=& Y(\hat{q}) - (B-L)(\hat{q}) \, , \\[12pt]
Y(\ell) - (B-L)(\ell) &=& Y(\hat{\ell}) - (B-L)(\hat{\ell}) \, ,
\end{array}
\label{hatanother}
\end{eqnarray}
with the additional constraints
\begin{eqnarray}
\begin{array}{rcl}
(B-L)(\hat{q}) &=& (B-L)(\ell) \, , \\[12pt]
(B-L)(\hat{\ell}) &=& (B-L)(q) \, ,
\end{array}
\label{hatBL}
\end{eqnarray}
so $B(\hat{q})=-1$ and $L(\hat{\ell})=-1/3$, as expected.

In this paper we pursue the possibility that the charge relations
in Eqs.~(\ref{newhyper}) and (\ref{another}), and the charge
dequantization described in Eq.~(\ref{hathyper}), are real, with
$B$ and $L$ as ungauged global symmetries, and quarks and leptons
as the ultimate constituents of ordinary matter. Thus the global,
robust against local interactions, fractional pieces of
hypercharge $4(3B-L)/3$ and $4(B-3L)/3$ as well as $B-L$ should
have a topological character originated by a hidden or broken
discrete electroweak symmetry between quarks and leptons, which
allows mixing between underlying local and topological integer
charges.  The idea that quarks and leptons may be entities with
topological structure has been discussed in the
literature,\cite{Rebbi,Vachas} but without addressing
Eqs.~(\ref{newhyper}) and (\ref{another}). In our previous
work,\cite{Matute1,Matute2} we have studied the weak topological
charge underlying quarks referring only to Eq.~(\ref{hyper}),
ignoring any charge structure defined in terms of $B$ and $L$, as
in Eq.~(\ref{hathyper}). Our aim is now to state the
principles/symmetries behind the described charge relations, show
the way they lead to such connections in the new scenario of
preleptons, and resolve the above puzzling questions.

The contents of this paper are organized as follows.  In Sec.
\ref{principles} we state the principles which provide the physics
rationale that sustain our known approach to the above charge
relationships between quarks and leptons.  The problem of anomaly
cancellation which appears because of the charge structure in
preleptons, but not in leptons, is discussed in Sec.
\ref{preleptons}. The mentioned questions prompted by the SM with
Dirac neutrinos are answered in Secs.
\ref{families}--\ref{confinement}. Our conclusions and outlook are
presented in Sec. \ref{conclusions}.

\section{The Principles}
\label{principles}

Compatible with the points made in Sec. \ref{introduction}, we
propose an extension of the SM with Dirac right-handed neutrinos
based on two principles.
\begin{arabiclist}
\item[(1)] Principle of electroweak quark--lepton symmetry: {\it
there exists a hidden discrete $Z_2$ symmetry in the electroweak
interactions of quarks and leptons}.
\end{arabiclist}
The basic and discrete symmetry between quarks and leptons
postulated in this principle is hidden in nature, broken in the
hypercharge sector according to the symmetric pattern of
Eqs.~(\ref{hyper}) and (\ref{newhyper}), and in the Yukawa sector
where there are strong differences of masses and mixing angles
which are not yet understood, though an universality can be
realized where the leading mass or mixing matrices are the same
for all quarks and leptons and small corrections can make the
whole difference.\cite{Smirnov}  Equations (\ref{hathyper}) and
(\ref{hatql}), with hypercharges specified in Eq.~(\ref{hyper}),
exhibit the integer charges underlying quarks and fractional
charges underlying leptons. These new charge states are being
referred to as prequarks and preleptons, respectively, and the
assumed discrete electroweak symmetry between prequarks and
leptons, and quarks and preleptons, which defines a $Z_2$ group,
is the so-called presymmetry.\cite{Matute2}  We say that
presymmetry is a dual symmetry in the sense that it operates for
both prequark--lepton and quark--prelepton systems.  We also note
that Majorana neutrinos are excluded from this picture because
combinations of baryon and lepton numbers appear as good global
symmetries in the robust fractional pieces of charge of Eqs.
(\ref{newhyper}) and (\ref{hathyper}) as well as in
Eq.~(\ref{another}).
\begin{arabiclist}
\item[(2)] Principle of weak topological-charge confinement: {\it
observable particles have no weak topological charge}.
\end{arabiclist}
This principle takes care of the fact that all weak topological
features have no observable effects at the zero-temperature scale
due to the smallness of the weak coupling. But it gives us the
necessary insights to discern the topological character of the
universal fractional piece of charge isolated in
Eqs.~(\ref{hyper}), (\ref{newhyper}) and (\ref{hathyper}). We
emphasize that this principle is secondary to that of gauge
confinement, in the sense that electroweak forces by themselves
cannot lead to actual confinement of topologically nontrivial
particles.  In the case of quarks, confinement is due to strong
color confinement, in spite of our basic assumption that they
appear as topologically nontrivial in the weak sector.

\begin{figure}
\begin{picture}(300,165)
\multiput(70,45)(190,0){2}{\circle{90}}
\multiput(70,45)(190,0){2}{\circle*{3}} \put(70,48){$\ell$}
\put(260,48){$\hat{\ell}$}
\multiput(93,60)(110,0){2}{\shortstack[l]{topological\\vacuum}}
\put(93,45){$n_{W}=0$} \put(93,25){$Q^{(3)}=0$}
\put(203,45){$n_{W}=4$} \put(203,25){$Q^{(3)}=1$}
\put(135,50){INSTANTON} \put(195,45){\vector(-1,0){60}}
\put(49,5){LEPTON}
\put(223,0){\shortstack[c]{TOPOLOGICAL\\PRELEPTON}}
\multiput(70,135)(190,0){2}{\circle{90}}
\multiput(70,135)(190,0){2}{\circle*{3}} \put(70,140){$\hat{q}$}
\put(260,140){$q$}
\multiput(93,150)(110,0){2}{\shortstack[l]{topological\\vacuum}}
\put(93,135){$n_{W}=0$} \put(93,115){$Q^{(3)}=0$}
\put(203,135){$n_{W}=4$} \put(203,115){$Q^{(3)}=1$}
\put(135,140){INSTANTON} \put(135,135){\vector(1,0){60}}
\put(42,95){PREQUARK}
\put(223,90){\shortstack[c]{TOPOLOGICAL\\QUARK}}
\end{picture}
\caption{Relations of presymmetric prequarks and leptons with
presymmetric topological quarks and topological preleptons. The
index $n_{W}$ denotes the topological winding number of the
associated weak gauge vacuum configuration.} \label{figure1}
\end{figure}
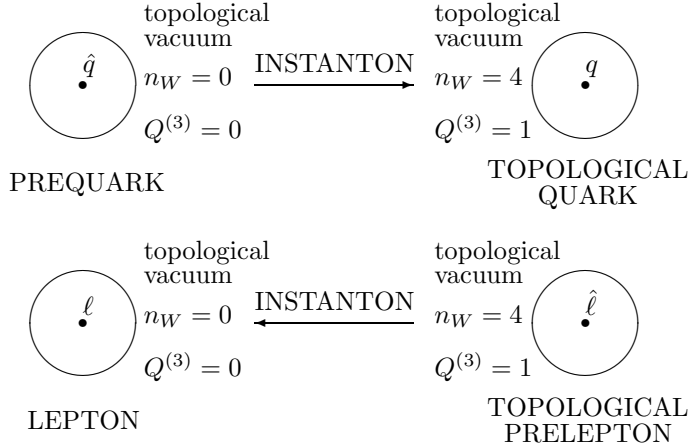

According to the principle of weak topological-charge confinement,
leptons and hadrons have associated zero weak topological charge,
in the sense of Eqs. (\ref{totalcharge})--(\ref{Z3}) set below. We
first distinguish between the fractionally-charged and
topologically-trivial constituent quarks of the SM and the
topological quarks of integer local charge and nonzero weak
topological charge of our model.\cite{Matute1,Matute2}  The
feature that in baryons quarks are confined in threes requires at
least a weak topological charge associated with a bookkeeping
$Z_3$ charge, defining a nontrivial value $+1$ for topological
quarks to have the equivalence between three topological quarks
and three constituent quarks. Hence, the~$3$ of this modulo charge
in topological quarks, based on counting, is due to the number of
colors. In the symmetric scenario of integer charged leptons and
prequarks with zero $Z_{3}$ charge, we shall see that the
connection between topologically-trivial prequarks and topological
quarks is a vacuum tunneling event --- a weak four-instanton ---
carrying positive topological charge, having associated a $+1$
modulo charge, and occurring between their topologically
inequivalent weak gauge vacuum configurations. In the dual case of
topological quarks and preleptons with fractional total charge,
preleptons also have, because of presymmetry, a $Z_{3}$ charge
equal to $+1$. The connection between topological preleptons and
leptons will be now a vacuum tunneling event, also a weak
four-instanton, associated with a negative topological charge and
a $-1$ $Z_3$-charge. These symmetries and relations are shown in
Fig.~\ref{figure1}. A similar mechanism operates for the change
from topological quarks to constituent quarks. We assume that the
weak topological vacua of leptons and constituent quarks are
equivalent.

The assumed charge structure underlying quarks and leptons is at
the end a mixing between the charges invoked by the two
principles: a local charge associated as usual with the local
field which creates the fermion and a topological charge related
to a weak instanton which cannot be generated by local
operations.\cite{Matute2}  This point is important and requires
more clarification. To make it we adapt the general
topological-charge arguments given by Wilczek for having
fractional charges.\cite{Wilczek}  Using the suffix $B$ to denote
bare charges, the renormalized charge with two underlying classes
of charge quantized in integers is defined by the master-charge
equation
\begin{equation}
Q_{R} = \mathit{z}^{(0)} \; Q_{B}^{(0)} + \mathit{z}^{(3)} \;
Q_{B}^{(3)} \, , \label{totalcharge}
\end{equation}
where the first charge is local and the second is topological, and
$\mathit{z}^{(0)}$ and $\mathit{z}^{(3)}$ are renormalization
factors. We note in passing that renormalization does not modify
charges themselves, but only the means by which they are measured,
i.e. couplings; therefore any charge structure in the elementary
fermions must have a topological character.  A conventionally
normalized charge for a general state is the ratio of its
renormalized charge and the minimal renormalized local charge. It
may be written as
\begin{equation}
Q_{N} = Q^{(0)} + \frac{\mathit{z}^{(3)}}{\mathit{z}^{(0)}} \;
Q^{(3)} \, , \label{shift}
\end{equation}
with an integer $Q^{(0)}$ and a ratio $\mathit{z}^{(3)} /
\mathit{z}^{(0)}$ that in a sense measures the induced charge
associated with a unit of topological charge.  But, if the
topological charge is integer defined modulo $3$, associated with
the additive $Z_3$ group, the renormalized charge spectrum for
$Q^{(3)}=3$ is the same as that for $Q^{(3)}=0$. This happens
because a topological charge $3$ configuration, being
topologically trivial, can be produced by local operations.  Then
the ratio $\mathit{z}^{(3)} / \mathit{z}^{(0)}$ between the units
of the renormalized topological and local charges must be
restricted to the form
\begin{eqnarray}
\displaystyle \frac{\mathit{z}^{(3)}}{\mathit{z}^{(0)}} =
\frac{n}{3} \, ,
\label{couplings}
\end{eqnarray}
\\[-9pt]
with an integer $n$.  Therefore the fractional piece of the
normalized charge which has a topological character will be always
a multiple of $1/3$.  Furthermore, for simplicity, it may be
introduced at a normalized {\it bare} level as
\begin{equation}
Q_{N} = Q^{(0)} + \frac{n}{3} \; Q^{(3)} \, , \label{Z3}
\end{equation}
which leads to a normalized fractional charge that can be treated
as an effective bare local charge at a lower level of description.
It is worth remarking that the modulo charge $Q^{(3)}$ coming from
the principle of weak topological-charge confinement is a new
index based on counting, different from the familiar winding and
Pontryagin topological numbers assigned to gauge field
configurations;\cite{book1} it will be specified in
Sec.~\ref{preleptons} according to the values of $B$ and $L$ of
the corresponding fermion.

We illustrate the meaning of Eq.~(\ref{Z3}) by considering the
normalized electric charge of quarks.  Using the topological
indexes $n=2$ and $Q^{(3)}=1$ (see Sec.~\ref{preleptons}) we
obtain, just as in Eq.~(\ref{Qcharge}),
\begin{eqnarray}
\begin{array}{rcl}
\displaystyle Q(u^{i}_{a}) &=& \displaystyle Q^{(0)}(u^{i}_{a}) +
\frac{2}{3} \, Q^{(3)}(u^{i}_{a}) = 0 + \frac{2}{3} = \frac{2}{3}
\, , \\[12pt] \displaystyle Q(d^{i}_{a}) &=& \displaystyle
Q^{(0)}(d^{i}_{a}) + \frac{2}{3} \, Q^{(3)}(d^{i}_{a}) = -1 +
\frac{2}{3} = - \frac{1}{3} \, ,
\end{array}
\label{Qquarks}
\end{eqnarray}
where $i$ denotes the color degree of freedom, $Q^{(0)}$ is
integer as in the lepton partner, and $Q^{(3)}$ is independent of
flavor. The $Z_3$ property of the $Q^{(3)}$ charge, with the $3$
related to color charge, means
\begin{equation}
Q(q^{i}_{a} q^{j}_{b} q^{k}_{c}) = Q^{(0)}(q^{i}_{a} q^{j}_{b}
q^{k}_{c}) + \frac{2}{3} \, Q^{(3)}(q^{i}_{a} q^{j}_{b} q^{k}_{c})
= Q^{(0)}(q^{i}_{a} q^{j}_{b} q^{k}_{c}) + 2 \, ,
\end{equation}
with a bare three-quark set, as in baryons, of $Q^{(3)}=3$ (the
neutral element of the additive $Z_3$ group) and two extra units
of normalized charge shared, being equivalent to an
integer-charged system of $Q^{(3)}=0$ constituted by three quarks
with fractional charge. The combination with strong color
confinement implies that the three color indexes are different
from each other.  At the lower level of description of the SM, the
fractional charge of constituents quarks is regarded as local,
encoding the conservation of topological charge as discussed
below.

All of these cannot be applied to leptons in the same way, because
they are colorless particles. We shall argue in
Sec.~\ref{families} that in this dual case $Z_3$ is associated
with the number of generations, so that instead of
Eq.~(\ref{Qquarks}) we have, using $Q^{(3)}=-1$,
\begin{eqnarray}
\begin{array}{rcl}
\displaystyle Q(\nu_{a}) &=& \displaystyle Q^{(0)}(\nu_{a}) +
\frac{2}{3} \, Q^{(3)}(\nu_{a}) = \frac{2}{3} - \frac{2}{3} = 0 \,
, \\[12pt] \displaystyle Q(e_{a}) &=& \displaystyle Q^{(0)}(e_{a})
+ \frac{2}{3} \, Q^{(3)}(e_{a}) = - \frac{1}{3} - \frac{2}{3} = -
1 \, ,
\end{array}
\label{Qleptons}
\end{eqnarray}
where now $Q^{(0)}$ is fractional as in the quark partner,
$Q^{(3)}$ is independent of flavor, and $Q$ is integer. This
implies
\begin{equation}
Q(\ell_{a} \ell_{b} \ell_{c}) = Q^{(0)}(\ell_{a} \ell_{b}
\ell_{c}) + \frac{2}{3} \, Q^{(3)}(\ell_{a} \ell_{b} \ell_{c}) =
Q^{(0)}(\ell_{a} \ell_{b} \ell_{c}) - 2 \, ,
\end{equation}
with three units of $Q^{(3)}$ charge making a trivial topology and
inducing an entire charge, though leptons themselves are integer
charged and topologically trivial, as seen from
Eqs.~(\ref{Qquarks}) and (\ref{Qleptons}) by invoking presymmetry,
and do not confine. Again, at the level of description of the SM,
the entire charge of leptons is simply regarded as local, ignoring
the cancellation of the fractional part of topological character.
Other insights are discussed in the next sections.

In Refs.~\refcite{Matute1} and \refcite{Matute2} we account for
the fractional charge of quarks relative to leptons, as in
Eqs.~(\ref{hyper}) and (\ref{Qcharge}), under the assumption that
only integral bare hypercharges are totally associated with local
fields. In the end the integer $n$ in Eq.~(\ref{Z3}) is identified
as a weak topological charge or Pontryagin index associated with
the infinite additive group $Z$, i.e. a genuine integer, not
modulo $3$.  We do start with quarks having an integer local bare
charge together with an exact electroweak discrete symmetry
between bare quarks and leptons, and end along the line exposed
above with quarks having fractional bare charges and a broken or
hidden symmetry. The special feature is that this breaking has a
timelike topological character related to a tunneling transition
from a boundary vacuum configuration of weak gauge fields at the
far past to a topologically different boundary vacuum
configuration at the far future --- \linebreak a weak
four-instanton. The finite universal 4/3 part of the quark
hypercharge indeed gets a topological meaning. The factor 4 is the
topological charge of the four-instanton fixed by the requirement
of gauge anomaly cancellation and the 3 is due to the color
number. It is important to note that the value of this topological
index is independent of the overall normalization of $Y$ in
Eq.~(\ref{charge}).  Also, the charge normalization to SM values
at the bare level, demanded by anomaly cancellation, and the zero
topological charge in hadrons effectively remove the extremely
large time scale associated with weak topological effects on
quarks, which is set by the smallness of the weak gauge coupling
in the phase of the zero-temperature scale of the model.

\section{Preleptons and Presymmetry}
\label{preleptons}

Here we apply the methodology developed in Refs.~\refcite{Matute1}
and \refcite{Matute2} to the new symmetric scenario of quarks and
preleptons, now including the bookkeeping charge from the additive
$Z_3$ group, and making clear the gauge-field solution associated
with topological preleptons and topological quarks. All of these
will provide the clues to answer the above pressing questions on
the SM with Dirac neutrinos. To make the point, we follow the
principle of electroweak quark--lepton symmetry and start out with
the dual scenario of constituent quarks and leptons in which the
fractional quark hypercharges are local and the integral
hypercharge of leptons relative to quarks relies just on the
universal $-$4/3 value, independent of flavor and handness, whose
topological character will be exposed from the requirement of
anomaly cancellation and the principle of weak topological-charge
confinement. In this case, Eqs. (\ref{totalcharge})--(\ref{Z3})
are used with a fractional $Q^{(0)}$ and an integer $Q$.  We
denote leptons with fractional bare charges, the so-called
preleptons, by hats over the symbols that represent the
corresponding leptons. Both scenarios of prequarks and preleptons
would be used by nature. The relevant points are the hidden
presymmetry and the topological character of the fractional charge
piece that normalizes the prequark and prelepton local charges to
the SM values.

It is assumed that each prelepton has the spin, weak isospin, and
flavor of the corresponding lepton.  Its bare weak hypercharge is
the same as in its quark partner specified in Eq.~(\ref{hyper})
and its bare electric charge is defined according to
Eq.~(\ref{charge}). Thus, in this scenario all local bare charges
in fermions are fractional.  The fractional lepton number of
preleptons can be fixed through the leptonic version of the
Gell-Mann--Nishijima formula for quarks:
\begin{equation}
Q = \frac{1}{2} \; (-L + L_{e} + L_{\nu_{e}} + L_{\mu} +
L_{\nu_{\mu}} + L_{\tau} + L_{\nu_{\tau}}) \, ,
\label{flavors}
\end{equation}
where $L_{e},\ldots,L_{\nu_{\tau}}$ denote (pre)lepton flavors.
This formula must hold for preleptons if we are to generate
leptons from them. The result $L(\hat{\ell})=-1/3$ agrees with
that obtained from Eqs.~(\ref{hatql}) and (\ref{hatBL}), giving
consistency to the whole picture. Tables \ref{table1} and
\ref{table2} list the quantum numbers which, together with the
spin $1/2$, define preleptons. From these, presymmetry and so the
electroweak discrete $Z_2$ symmetry, which relate bare quarks and
preleptons, are easily perceived. Again, we note that the baryon
minus lepton number $B-L$ is the meaningful quantum number to be
consider under presymmetry between quarks and preleptons.
\begin{table}[t]
 \tbl{\label{table1} Prelepton additive quantum numbers.}
  {\begin{tabular}{@{}lrrrrrr@{}} \toprule
  \hspace{1.1cm} Prelepton & \hspace{0.1cm} $\hat{\nu}_{e}$ &
  \hspace{0.1cm} $\hat{e}$ & \hspace{0.1cm}
  $\hat{\nu}_{\mu}$ & \hspace{0.1cm} $\hat{\mu}$ & \hspace{0.1cm}
  $\hat{\nu}_{\tau}$ & \hspace{0.1cm}
  $\hat{\tau}$ \\  \colrule \\[-7pt]
  $L$ (lepton number) & \hspace{0.1cm} $\displaystyle - \frac{1}{3}$ &
  \hspace{0.1cm} $\displaystyle - \frac{1}{3}$ & \hspace{0.1cm}
  $\displaystyle - \frac{1}{3}$ &   \hspace{0.1cm} $\displaystyle
  - \frac{1}{3}$ & \hspace{0.1cm} $\displaystyle - \frac{1}{3}$ &
  \hspace{0.1cm} $\displaystyle - \frac{1}{3}$ \\[10pt]
  $Q$ (electric charge) & \hspace{0.1cm} $\displaystyle \frac{2}{3}$ &
  \hspace{0.1cm} $\displaystyle - \frac{1}{3}$ & \hspace{0.1cm}
  $\displaystyle \frac{2}{3}$ &   \hspace{0.1cm} $\displaystyle
  - \frac{1}{3}$ & \hspace{0.1cm} $\displaystyle \frac{2}{3}$ &
  \hspace{0.1cm} $\displaystyle - \frac{1}{3}$ \\[10pt]
  $L_{\nu_{e}}$ (electron neutrinoness) & 1 & 0 & 0 & 0 & 0 & 0
  \\[3pt]
  $L_{e}$ (electroness) & 0 & $-$1 & 0 & 0 & 0 & 0 \\[3pt]
  $L_{\nu_{\mu}}$ (muon neutrinoness) & 0 & 0 & 1 & 0 & 0 & 0
  \\[3pt]
  $L_{\mu}$ (muoness) & 0 & 0 & 0 & $-$1 & 0 & 0 \\[3pt]
  $L_{\nu_{\tau}}$ (tau neutrinoness) & 0 & 0 & 0 & 0 & 1 & 0
  \\[3pt]
  $L_{\tau}$ (tauness) & 0 & 0 & 0 & 0 & 0 & $-$1 \\[2pt]
  \botrule
  \end{tabular}}
  \end{table}
\begin{table}
  \tbl{\label{table2} Weak isospin and hypercharge of preleptons.}
  {\begin{tabular}{@{}ccrcr@{}} \toprule
  Prelepton & \hspace{0.5cm} $\hat{\nu}_{aL}$ & \hspace{0.5cm}
  $\hat{e}_{aL}$ & \hspace{0.5cm} $\hat{\nu}_{aR}$
  & \hspace{0.5cm} $\hat{e}_{aR}$  \\  \colrule \\[-7pt]
  $T$ & \hspace{0.5cm} $\displaystyle \frac{1}{2}$ & \hspace{0.5cm}
  $\displaystyle \frac{1}{2}$   & \hspace{0.5cm} 0 & \hspace{0.5cm} 0
  \\[10pt]   $T_{3}$ & \hspace{0.5cm} $\displaystyle \frac{1}{2}$ &
  \hspace{0.5cm}   $\displaystyle - \frac{1}{2}$ & \hspace{0.5cm} 0
  & \hspace{0.5cm} 0 \\[10pt]   $Y$ & \hspace{0.5cm} $\displaystyle
  \frac{1}{3}$ &   \hspace{0.5cm} $\displaystyle \frac{1}{3}$
  & \hspace{0.5cm} $\displaystyle \frac{4}{3}$ & \hspace{0.5cm}
  $\displaystyle - \frac{2}{3}$ \\[7pt] \botrule
 \end{tabular}}
 \end{table}

At the classical level, we take the point of view that what fields
with and how preleptons (and prequarks) interact are dictated by
the SM,\cite{book1} with a Lagrangian that looks the same as that
of quarks and leptons excepting hypercharge couplings and
incorporation of Dirac right-handed neutrinos. Analytically,
presymmetry is then invariance of the bare electroweak Lagrangian
for quarks and preleptons under the flavor transformation
\begin{equation}
u^{i}_{aL} \leftrightarrow \hat{\nu}_{aL} \, , \qquad u^{i}_{aR}
\leftrightarrow \hat{\nu}_{aR} \, , \qquad d^{i}_{aL}
\leftrightarrow \hat{e}_{aL} \, , \qquad d^{i}_{aR}
\leftrightarrow \hat{e}_{aR} \, , \label{presy}
\end{equation}
\\[-9pt]
with no change for the standard gauge and Higgs fields.

The fractional hypercharge of preleptons leads, however, to the
appearance of triangle gauge anomalies at the quantum level. In
Ref.~\refcite{Matute2} these were eliminated in the context of
prequarks by adding Chern--Simons counterterms to the Lagrangian
of the system. We now work out this problem in the scenario of
preleptons and in light of the principle of weak
topological-charge confinement, which introduces a bookkeeping
charge modulo~$3$. Although this part reproduces more or less the
contents of our previous work, we consider it necessary to clarify
the key topological arguments of our approach with the new
insights gained in the present formulation.

Following standard literature,\cite{book1} it is found that the
$\mbox{U(1)}_{Y}$ gauge current of constituent quarks and
preleptons
\begin{equation}
\hat{J}^{\mu}_{Y} = \overline{q}_{aL} \gamma^{\mu} \frac{Y}{2}
q_{aL} + \overline{q}_{aR} \gamma^{\mu} \frac{Y}{2} q_{aR} +
\overline{\hat{\ell}}_{aL} \gamma^{\mu} \frac{Y}{2}
\hat{\ell}_{aL} + \overline{\hat{\ell}}_{aR} \gamma^{\mu}
\frac{Y}{2} \hat{\ell}_{aR} \, ,
\label{gaugecurrent}
\end{equation}
exhibits the $\mbox{U(1)}_{Y} [\mbox{SU(2)}_{L}]^{2}$ and
$[\mbox{U(1)}_{Y}]^{3}$ anomalies:
\begin{eqnarray}
\partial_{\mu} \hat{J}^{\mu}_{Y} & = & - \frac{g^{2}}{32 \pi^{2}}
\left( \, \sum_{q_{L} , \hat{\ell}_{L}} \frac{Y}{2} \right)
\mbox{tr} \; W_{\mu\nu} \tilde{W}^{\mu\nu} \nonumber \\
& &  \nonumber \\
& & \, - \frac{{g'}^{2}}{48 \pi^{2}} \left( \, \sum_{q_{L} ,
\hat{\ell}_{L}} \frac{Y^3}{2^3} - \sum_{q_{R} , \hat{\ell}_{R}}
\frac{Y^3}{2^3} \right) F_{\mu\nu} \tilde{F}^{\mu\nu} \, , \;\;
\label{anomaly1}
\end{eqnarray}
where $g$, $g'$ and $W_{\mu\nu}$, $F_{\mu\nu}$ are the
$\mbox{SU(2)}_{L}$, $\mbox{U(1)}_{Y}$ couplings and field
strengths, respectively, $\tilde{W}^{\mu\nu}$ and similarly
$\tilde{F}^{\mu\nu}$ is defined by
\begin{equation}
\tilde{W}_{\mu\nu} = \frac{1}{2} \; \epsilon_{\mu\nu\lambda\rho}
\; W^{\lambda\rho} \, ,
\label{Wtilde}
\end{equation}
and the hypercharge sums run over all the fermion representations.
The anomalies appear because these sums do not vanish.

The terms on the right-hand side in Eq.~(\ref{anomaly1}) are
divergences of gauge-dependent currents which can be written in
the form
\begin{equation}
\partial_{\mu} \hat{J}^{\mu}_{Y} = - N_{\hat{\ell}} \; \partial_{\mu}
J^{\mu}_{T} \, ,
\label{divJ}
\end{equation}
where $N_{\hat{\ell}} = 4 N_{g}$ is the number of left- and
right-handed preleptons, $N_{g}$ denotes the number of
generations, and
\begin{equation}
J^{\mu}_{T} = \frac{1}{4 N_{\hat{\ell}}} \, K^{\mu} \sum_{q_{L} ,
\hat{\ell}_{L}} Y + \frac{1}{16 N_{\hat{\ell}}} \, L^{\mu} \left(
\, \sum_{q_{L} , \hat{\ell}_{L}} Y^{3} - \sum_{q_{R} ,
\hat{\ell}_{R}} Y^{3} \right) = \frac{1}{6} \, K^{\mu} -
\frac{1}{8} \, L^{\mu} \, . \label{currX}
\end{equation}
\\[1pt]
It is worth noting that the last equality in Eq.~(\ref{currX}) is
independent of the overall normalization of $Y$ in
Eq.~(\ref{charge}). The $K^{\mu}$ and $L^{\mu}$ are the
topological currents or Chern--Simons classes related to the
$\mbox{SU(2)}_{L}$ and $\mbox{U(1)}_{Y}$ gauge groups,
respectively. They are given by
\\[-6pt]
\begin{eqnarray}
\begin{array}{rcl}
\displaystyle K^{\mu} &=& \displaystyle \frac{g^{2}}{8 \pi^{2}} \,
\epsilon^{\mu\nu\lambda\rho} \; \mbox{tr} \!\! \left( W_{\nu}
\partial_{\lambda} W_{\rho} - \frac{2}{3} \, i g
W_{\nu} W_{\lambda} W_{\rho} \right) , \\[12pt]
\displaystyle L^{\mu} &=& \displaystyle \frac{{g'}^{2}}{12
\pi^{2}} \, \epsilon^{\mu\nu\lambda\rho} A_{\nu}
\partial_{\lambda} A_{\rho} \, .
\end{array}
\label{CS}
\end{eqnarray}
\\[-8pt]

A new current can now be defined from Eq.~(\ref{divJ}):
\begin{equation}
J^{\mu}_{Y} = \hat{J}^{\mu}_{Y} + N_{\hat{\ell}} \; J^{\mu}_{T} \,
,
\label{newJ}
\end{equation}
which is conserved but it is gauge-dependent.  The presymmetric
local counterterm to be added to the bare Lagrangian of quarks and
preleptons, needed to obtain this anomaly-free current, is
\begin{equation}
\Delta {\cal L} = g' N_{\hat{\ell}} \, J^{\mu}_{T} A_{\mu} \, ,
\label{Lagran}
\end{equation}
\\[-8pt]
so that, due to the antisymmetry of
$\epsilon^{\mu\nu\lambda\rho}$, only the non-Abelian fields will
be topologically relevant, as expected.  It is also gauge
noninvariant.  However, it has nontrivial topological attributes
to be described in the following, which allow to solve all
problems at once.  In fact, presymmetry has available a degree of
freedom which can be chosen conveniently through this counterterm
to restore gauge invariance.  Certainly, a fixed nonzero value
breaks such a symmetry.  The parameter we are referring to is the
topological charge or Pontryagin index.

Actually, the charge corresponding to the current in
Eq.~(\ref{newJ}) is not conserved because of the existence of the
topological charge associated with gauge fields.  To see this, the
charge change between $t = - \infty$ and $t = + \infty$ is
calculated, concentrating on Euclidean space--time solutions which
have well-defined integer topological charges.\cite{Farhi}  This
is because in our approach the prelepton--lepton and
prequark--quark transitions are in the end vacuum tunneling events
in Minkowski space--time at the zero-temperature scale --- weak
instantons. As usual, it is assumed that the region of space--time
where the energy density is nonzero is bounded. Therefore this
region can be surrounded by a three-dimensional surface with the
vacuum at this boundary on which the field configuration becomes
pure gauge, i.e. $W_{\mu} = - (i/g) (\partial_{\mu} U) U^{-1}$
where the transformation $U$ takes values in the corresponding
gauge group. In this case, using Eqs.~(\ref{currX})--(\ref{newJ})
just for the pure gauge fields, such a charge becomes,
independently of the overall normalization of $Y$ in
Eq.~(\ref{charge}),
\begin{equation}
Q_{Y}(t) = \int d^{3}x \; J^{o}_{Y} = \frac{N_{\hat{\ell}}}{6} \;
n_{W}(t) \, ,
\label{nQ3}
\end{equation}
where
\begin{equation}
n_{W}(t) = \frac{1}{24 \pi^2} \int d^{3}x \, \epsilon^{ijk} \,
\mbox{tr} (\partial_{i}U U^{-1}
\partial_{j}U U^{-1} \partial_{k}U U^{-1})
\label{winding}
\end{equation}
is the topological winding number of the $\mbox{SU(2)}_{L}$ gauge
transformation.  This number is integer-valued if we consider a
fixed time $t$ and assume that $U \rightarrow 1$ for $|
\mbox{\boldmath$x$}| \rightarrow \infty$.  This $U$ is then a map
from an equal time sphere $S^{3}$ onto the sphere of parameters
$S^{3}$ of the $\mbox{SU(2)}_{L}$ group manifold. These maps are
classified according to the homotopy classes which are labelled by
an integer topological index.  The integral given by
Eq.~(\ref{winding}) computes this integer.  For the Abelian case,
$n_{W}=0$.

Consequently, for the gauge field configurations associated with
preleptons which at the initial $t = - \infty$ and the final $t =
+ \infty$ are supposed of the above pure gauge form at the
boundary with different winding numbers, the change in charge is
\\[-6pt]
\begin{equation}
\Delta Q_{Y} = \displaystyle \frac{N_{\hat{\ell}}}{6} \left[
n_{W}(t=+\infty) - n_{W}(t=-\infty) \right] .
\label{deltaQ1}
\end{equation}
\\[-8pt]
This difference between the integral winding numbers can be
represented by the topological charge defined by
\\[-7pt]
\begin{equation}
Q_{T} = \int d^{4}x \, \partial_{\mu} K^{\mu} = \frac{g^{2}}{16
\pi^{2}} \int d^{4}x \, \mbox{tr} (W_{\mu\nu} \tilde{W}^{\mu\nu})
\, , \label{QT1}
\end{equation}
\\[-7pt]
where the integration is over all of (Euclidean) space--time,
assuming that $K^{i}$ decreases rapidly enough at spatial
infinity.  In our Euclidean approach with pure gauge
configurations, this topological index is gauge invariant,
conserved and integer-valued. Then we have
\begin{equation}
Q_{T} = Q^{(3)}n = n_{W}(t=+\infty) - n_{W}(t=-\infty) \, ,
\label{QT2}
\end{equation}
so producing an integer winding number change. Using gauge freedom
we can set $n_{W}(-\infty)=0$ or $n_{W}(+\infty)=0$ and interpret
the topological charge as the winding number of the gauge
configuration to which the field tends or starts from. Assuming
these values and the above principles, together with the
topological charge we have introduced the $Z_3$ counting number
$Q^{(3)}$, equal to $\pm 1$ for nontrivial topology and $0$ for
trivial topology, just as if the topological number were itself a
proper $Z_3$ charge, whose crucial modulo property is exclusively
due to those principles, although leptons do not confine.
Therefore the change in winding number is also determined by the
charge defined modulo $3$.

We emphasize that the integral topological charge in
Eq.~(\ref{QT1}) is associated with an Euclidean field
configuration that, according to Eq.~(\ref{QT2}), interpolates in
imaginary time between the real time pure-gauge configurations in
the far past and future which have different topological winding
numbers.  The condition is the one established for instantons.
Thus the continuous interpolation is considered as a tunneling
process in Minkowski space--time, where all fields (fermions and
bosons) really reside, so that a barrier must separate the initial
and final vacuum gauge field configurations. If the analytical
continuation to Euclidean space--time is not made, a nonzero
topological charge implies a nonvanishing energy density at
intermediate real time and then no conservation of energy.

Equation~(\ref{deltaQ1}) can then be written as
\begin{equation}
\Delta Q_{Y} = N_{\hat{\ell}} \; \frac{n}{6} \; Q^{(3)} \, .
\label{deltaQ2}
\end{equation}
A consequence of this result is that, because of the global
character of the topological index $n$ and the modulo charge
$Q^{(3)}$, $N_{\hat{\ell}}$ preleptons have to change their $Y/2$
in the same amount $nQ^{(3)}/6$. For each prelepton it implies the
hypercharge shift
\begin{equation}
Y(\hat{\ell}) \rightarrow Y(\hat{\ell}) + \frac{n}{3} \;
Q^{(3)}(\hat{\ell}) \, ,
\label{norma}
\end{equation}
hence giving an extra contribution which changes its initial
fractional bare value, just as if the $Z_{3}$ charge were a proper
topological charge and the index $n$ were an integer to be fixed
from the condition of anomaly cancellation. So the charge in
preleptons does not only depend on fields at a single instant of
time, but also on flow of current at infinity.  On the other hand,
to make sense of this bare charge normalization we invoke
Eqs.~(\ref{totalcharge})--(\ref{Z3}) with a fractional $Q^{(0)}$
just as in the topological quark partner, defining $Q^{(3)}=-1$
for the charge added to topological preleptons.  This is done
because $Q^{(3)}=1$ in presymmetric topological quarks and
preleptons, whereas leptons have $Q^{(3)}=0$ and do not confine
(see Fig.~\ref{figure1}).

From Eq.~(\ref{hathyper}), it is natural to associate the
$Q^{(3)}$ attached to preleptons (prequarks) with the following
combination of $B$ and $L$:
\begin{equation}
Q^{(3)} = -(B-3L) \, ,
\label{Q3charge}
\end{equation}
noting that three preleptons (prequarks) make a system of
$Q^{(3)}=-3 (+3)$, the neutral element of the additive group
$Z_{3}$, though leptons do not confine.

Now the gauge anomalies must be recalculated. As in the dual
scenario of prequarks,\cite{Matute1,Matute2} it is found that they
are cancelled for $n=4$, independently of the overall
normalization of $Y$, because
\begin{eqnarray}
\sum_{q_{L}, \ell_{L}} Y = 0 \, , \qquad  \displaystyle
\sum_{q_{L} , \ell_{L}} Y^{3} - \sum_{q_{R} , \ell_{R}} Y^{3} = 0
\, .
\label{sumsY}
\end{eqnarray}

Thus, from Eq. (\ref{QT2}) with $Q^{(3)}=-1$ and $n=4$, we
conclude that the symmetric topological preleptons and topological
quarks have associated vacuum gauge field configurations of
winding number $n_{W}=4$, if we use gauge freedom to set $n_{W}=0$
for the topological vacuum in leptons and constituent quarks. The
prelepton--lepton transformation is then a vacuum tunneling event,
corresponding to a $Q_{T}=-4$ weak instanton (see
Fig.~\ref{figure1}). Similarly, in the prequark--lepton scenario,
the transition from prequarks with $n_{W}=0$ vacuum configuration
to topological quarks with $n_{W}=4$, and next to constituent
quarks having $n_{W}=0$, is via $Q_{T}=4$ and $Q_{T}=-4$ weak
instantons, respectively, with three topological quarks being
equivalent to three constituent quarks.

The hypercharge normalization in Eq.~(\ref{norma}) with
$Q^{(3)}=-1$ and $n=4$ agrees with Eq.~(\ref{hathyper}), and so
with Eqs.~(\ref{hyper}) and (\ref{newhyper}).  From
Eqs.~(\ref{charge}) and (\ref{flavors}), this also implies shifts
in prelepton bare charges equal to
\begin{eqnarray}
\begin{array}{rcl}
\Delta Q &=& \displaystyle \frac{1}{2} \, \Delta Y = - \frac{2}{3}
\, (B-3L) = - \frac{2}{3} \, , \\[12pt] \Delta L &=& \displaystyle
- 2 \, \Delta Q = \frac{4}{3} \, (B-3L) = \frac{4}{3} \, .
\end{array}
\end{eqnarray}
The addition of these universal fractional charges to the
fractional quantum numbers of preleptons gives the integer bare
charges of leptons.

It is worth remarking that the gauge current of fractional-charged
quarks and preleptons in Eq. (\ref{gaugecurrent}) is indeed an
effective local current in the sense that these particles have an
underlying charge structure according to presymmetry and the dual
prequark--lepton description based on integer local charges. Such
a charge structure in topological preleptons disappears from
leptons because of the weak four-instanton effect.

The above charge normalization leads to restoration of gauge
invariance, breaking of presymmetry in the Abelian sector,
dressing of preleptons into integer-charged bare leptons, and the
substitution of the bare presymmetric model by the SM with Dirac
neutrinos. In fact, from Eqs.~(\ref{currX}) and (\ref{sumsY}) we
note that the gauge-dependent topological current $J^{\mu}_{T}$ is
cancelled. But, the corresponding topological charge is gauge
invariant and manifests itself as a universal part of the
prelepton hypercharge according to Eq.~(\ref{hathyper}).  More
precisely, if we consider the bare topological current
$N_{\hat{\ell}} J^{\mu}_{T}$ in Eq.~(\ref{Lagran}) which causes
the hypercharge obtained in Eq.~(\ref{deltaQ2}), we can define an
effective prelepton local current
$\hat{J}^{\mu}_{Y,\mbox{\footnotesize eff}}$ by
\begin{equation}
\displaystyle \hat{J}^{\mu}_{Y,\mbox{\footnotesize eff}} = -
\frac{2}{3} \left( \overline{\hat{\ell}}_{aL} \gamma^{\mu} (B-3L)
\hat{\ell}_{aL} + \overline{\hat{\ell}}_{aR} \gamma^{\mu} (B-3L)
\hat{\ell}_{aR} \right)
\end{equation}
to absorb their effects, i.e. gauge anomaly cancellation, trivial
topology, and induced charge $- 2(B-3L)/3$ on preleptons
regardless of flavor, handness, and overall normalization of $Y$
in Eq.~(\ref{charge}). Thus Eq.~(\ref{newJ}) takes the
gauge-independent form
\begin{eqnarray}
J^{\mu}_{Y} & = & \overline{q}_{aL} \gamma^{\mu} \frac{Y}{2}
q_{aL} + \overline{q}_{aR} \gamma^{\mu} \frac{Y}{2} q_{aR} +
\overline{\hat{\ell}}_{aL} \gamma^{\mu}
\frac{Y-4(B-3L)/3}{2} \hat{\ell}_{aL} \nonumber \\ & & \nonumber \\
& & + \overline{\hat{\ell}}_{aR} \gamma^{\mu}
\frac{Y-4(B-3L)/3}{2} \hat{\ell}_{aR} \, .
\end{eqnarray}
At this point, preleptons with integral local hypercharge, which
effectively includes the finite universal $-4(B-3L)/3$ part, have
to be identified with the topologically trivial standard bare
leptons, which are the ones to start with in the usual quantum
field theory analysis (see Eq.~(\ref{Z3})); after this bare charge
normalization, quantum anomalies are absent to all orders of
perturbation theory. The replacement of preleptons by leptons in
the weak and Yukawa sectors is straightforward as they have the
same flavors and weak isospins. On the other hand, the explicit
charge normalization to standard values at the bare level,
required by anomaly cancellation, and the zero topological charge
in leptons effectively get rid of the extremely large time scale
associated with weak topological effects on preleptons, occurring
in general if the gauge fields are assumed to remain in the small
coupling regime, i.e. in the phase of the zero-temperature scale
of the model.\cite{tHooft}  Of course, the above results obtained
in the scenario of quarks and preleptons, somewhat duplicate those
found in the dual scenario of prequarks and
leptons.\cite{Matute1,Matute2}

\section{Triplication of Fermion Families}
\label{families}

In the framework of bare leptons with integer local charges, the
number~$3$ by which the index $n=4$ is divided in
Eq.~(\ref{hyper}), i.e. the order of the additive group $Z_{3}$
associated with the topological charge in bare quarks, corresponds
to the color number $N_{c}$.\cite{Matute2}  This is also seen from
Eq.~(\ref{newhyper}) since each quark has a baryon number fixed by
its number of colors:
\begin{equation}
\displaystyle B(q) = \frac{1}{N_{c}} \, .
\label{Bquark}
\end{equation}
However, in the present dual scenario of local quark charges,
the~3 cannot be interpreted in the same manner because color is
not a quantum number of leptons. Clearly, quarks and leptons have
to be treated differently in the explanation of the number~3 in
the additive group $Z_3$. In the lepton sector, it must be
associated with a numerable property of leptons and the number of
families $N^{(\ell)}_{g}$ is the only available degree of freedom
that can take the value~3. In other words, whereas the partition
of topological charges in the scenario with prequarks is according
to the number of colors, the partition in the scenario with
preleptons must be according to the number of generations.
Therefore to relate the~3 in Eq.~(\ref{norma}), i.e. the order of
the additive group $Z_3$ for the topological charge in colorless
preleptons, with the number of lepton families is unavoidable. It
would be surprising that this number~3 in $Z_3$ does not be
connected to the number of generations, considering the fact that
the SM offers no explanation for the existence of three
generations. From our point of view, this explanation comes in
terms of our two principles if we make the assumption that the
partition of topological charges must be explained by physics of
the SM: first, the family replication is to provide the degree of
freedom by means of which the required partition of topological
charges in the colorless lepton sector, according to the principle
of weak topological-charge confinement, is possible. Second, the
number of families is fixed by the presymmetry introduced with the
principle of electroweak quark--lepton symmetry, which relates the
partition of topological charges in the lepton and quark sectors.
Thus, presymmetry gives a connection between the number of
generations and the number of colors, and so the reason for the
existence of just three generations. This correlation takes place
because of the above assumption on the partition of topological
charges. Otherwise, there might be as well more physics which
could explain the number 3 in the additive group $Z_3$ of the
lepton sector, leaving no argument for its connection with the
number of generations.

The factor in realizing this approach is simply and directly that
each prelepton has a lepton number defined by its number of
families:
\begin{equation}
\displaystyle L(\hat{\ell}) = - \frac{1}{N^{(\ell)}_{g}} \, .
\label{Lprelepton}
\end{equation}
Therefore, while in the case of presymmetric prequarks and leptons
we have
\begin{equation}
(B-L)(\hat{q}) = (B-L)(\ell) = -1 \, ,
\end{equation}
in the case of preleptons and quarks presymmetry implies
\begin{equation}
(B-L)(\hat{\ell}) = \frac{1}{N^{(\ell)}_{g}} = (B-L)(q) =
\frac{1}{N_{c}} = \frac{1}{3} \, ,
\end{equation}
so that $N^{(\ell)}_{g}=N_{c}=3$.

On the other hand, if Eq.~(\ref{hathyper}) is used to have the
hypercharge of each prelepton and then the coefficient of
$K^{\mu}$ in Eq.~(\ref{currX}) is calculated, we end up with the
hypercharge shift
\begin{equation}
\displaystyle \Delta Y(\hat{\ell}) = - n \, Q^{(3)} L(\hat{\ell})
= \frac{n}{N^{(\ell)}_{g}} \; Q^{(3)}(\hat{\ell})
\label{deltaYl}
\end{equation}
in Eq.~(\ref{norma}). This leads to the result $\Delta
Y(\hat{\ell})=-4/3$ if $Q^{(3)}=-1$. Similarly, in the scenario of
prequarks, it can be shown that\cite{Matute2}
\begin{equation}
\displaystyle \Delta Y(\hat{q}) = - \frac{n}{N_{c}} \; Q^{(3)}
B(\hat{q}) = \frac{n}{N_{c}} \; Q^{(3)}(\hat{q}) \, ,
\label{deltaYq}
\end{equation}
and the result $\Delta Y(\hat{q})=4/3$ if $Q^{(3)}=1$. Equations
(\ref{deltaYl}) and (\ref{deltaYq}) exhibit the required meaning
of the $3$ in the hypercharge relationships of Eqs.~(\ref{hyper}),
(\ref{newhyper}) and (\ref{hathyper}), which validate
Eq.~(\ref{Lprelepton}). The general rule for the modulo charge of
the four-instanton configuration for preleptons and prequarks has
been given in Eq.~(\ref{Q3charge}).

In a sense, our solution of the fermion family problem involves a
quark (charge) structure and a (hidden) quark--lepton family
symmetry. The key point turns out to be presymmetry which in the
end requires $N^{(\ell)}_{g}=N^{(q)}_{g}$ and
$N^{(\ell)}_{g}=N_{c}$, correlating the number of quark--lepton
generations with the number of quark colors as put forward in
Ref.~\refcite{Matute3}; this correlation is not really surprising
in models that advocate a non-Abelian duality in quark--lepton
physics.\cite{HongMo}

\section{Topological Charge and Conservation of $B-L$}
\label{topcharge}

The underlying topologically nontrivial gauge field configuration
in bare leptons suggests to associate a bookkeeping charge with
leptonic matter.  If we consider that an effective fractional
topological charge equal to $Q_{T} = nQ^{(3)}/N_{\ell} = -1/N_{g}$
(using $n=4$ and $Q^{(3)}=-1$) can be associated with each bare
lepton, the following rule for the one associated with leptonic
matter is obtained now taking $Q^{(3)}=3B-L$ from
Eq.~(\ref{hatql}): $Q_{T} = - L/N_{g}$. Moreover, taking into
account that in the dual scenario the effective topological charge
in baryons is $Q_{T} = nQ^{(3)}/N_{q} = B/N_{g}$,\cite{Matute2}
the general result appears to be
\begin{equation}
Q_{T} = \frac{B-L}{N_{g}} \, .
\label{B-L}
\end{equation}

We see that the topological charge labels topological preleptons
and topological quarks.  At the lepton and quark bare level of the
SM, it is encoded into the bookkeeping charge defined in
Eq.~(\ref{B-L}). Also, conservation of the effective topological
charge in local interactions implies $B-L$ conservation, which in
turn forbids neutrino mass terms of Majorana type.   This outcome
is consistent with presymmetry which requires Dirac neutrinos.

The consistency of the definition of the bookkeeping charge in
Eq.~(\ref{B-L}) can be seen, for example, in baryon plus lepton
number ($B+L$) nonconservation processes associated with axial
anomalies in the field of weak instantons.\cite{tHooft}  It is
sufficient to recall that for three generations, one weak
instanton characterized by a topological charge $Q_{T}=1$, not
modulo $3$, forces via the anomaly baryon and lepton number
violations in three units: $\Delta B = \Delta L = -3$, matching
three baryons or nine quarks and three antileptons. A decay such
as $p+n+n \rightarrow e^{+}+ \bar{\nu}_{\mu}+\bar{\nu}_{\tau}$ is
then allowed, with an extremely large lifetime (exceeding the
lifetime of the Universe) typical of weak topological effects in
the zero-temperature phase. But, three antileptons make precisely
the same topological charge of the weak instanton.  Thus the
definition in Eq.~(\ref{B-L}) brings back topological charge
conservation in quantum flavor dynamics.  In fact, in Minkowski
space--time one weak instanton corresponds to a process in which a
topological charge $Q_{T} = 1$ is carried out.  It induces a
process with effective topological charge $Q_{T} = 1$ associated
with the generation of three antileptons. Similarly, the three
baryons make an effective topological charge $Q_{T} = 1$, so that
the bookkeeping charge defined in Eq.~(\ref{B-L}) is conserved.

Next we show that there is also a direct link between $B-L$ and
the charge modulo $3$ based on counting, which is apparent from
Eqs.~(\ref{hatanother}), (\ref{deltaYl}) and (\ref{deltaYq}). By
taking into account the principle of weak topological-charge
confinement and presymmetry we are led to define the following
$Z_{3}$ charges for (pre)quarks and (pre)leptons:
\begin{equation}
Q^{(3)}(q) = 1 \, , \qquad Q^{(3)}(\ell) = 0 \, , \qquad
Q^{(3)}(\hat{q}) = 0 \, , \qquad Q^{(3)}(\hat{\ell}) = 1 \, .
\label{topnum}
\end{equation}
The weak field boundary in prefermions and fermions are in
different homotopy classes and connected by the weak
four-instanton associated with $Q^{(3)}=1$ for prequarks and
$Q^{(3)}=-1$ for preleptons, as defined by Eq.~(\ref{Q3charge}).
Its addition to the $Z_{3}$ charge of prequarks and preleptons
gives the correct charge of quarks and leptons, respectively (see
Fig.~\ref{figure1}). Consequent with Eqs.~(\ref{Bquark}),
(\ref{Lprelepton}) and (\ref{topnum}), we find the relation
\begin{equation}
Q^{(3)} = N_{g} (B - L) \, , \label{BL}
\end{equation}
which provides us with a rule to assigning modulo charge to
matter, given the ones associated with (pre)quarks and
(pre)leptons. The conservation of $Q^{(3)}$ in local interactions
is the corresponding conservation of $B-L$. It is notable that
Eqs.~(\ref{Q3charge}), (\ref{deltaYl}), (\ref{deltaYq}) associated
with the weak four-instanton and (\ref{BL}) related to
(pre)fermions themselves be consistent with
Eq.~(\ref{hatanother}). By combining Eqs.~(\ref{B-L}) and
(\ref{BL}) we also have
\begin{equation}
Q^{(3)} = N_{g}^{2} Q_{T} \, .
\end{equation}

\section{Charge Quantization and Ungauging of $B-L$}
\label{Qquantization}

We now describe the way our approach leads to charge quantization.
It is known that charge becomes dequantized if the right-handed
neutrinos are included in the SM.  In fact, constraints of anomaly
cancellation and nonvanishing of all Dirac fermion masses lead to
hypercharges which depend on an arbitrary parameter $\epsilon$
as\cite{Foot}
\begin{eqnarray}
\begin{array}{rcll}
\displaystyle Y(u_{aL}) &=& \displaystyle \frac{1}{3} -
\frac{\epsilon}{3} \, ,
& \qquad Y(\nu_{aL}) = -1 + \epsilon \, , \\[12pt]
\displaystyle Y(d_{aL}) &=& \displaystyle \frac{1}{3} -
\frac{\epsilon}{3} \, ,
& \qquad Y(e_{aL}) = -1 + \epsilon \, , \\[12pt]
\displaystyle Y(u_{aR}) &=& \displaystyle \frac{4}{3} -
\frac{\epsilon}{3} \, ,
& \qquad Y(\nu_{aR}) = \epsilon \, , \\[12pt]
\displaystyle Y(d_{aR}) &=& \displaystyle - \frac{2}{3} -
\frac{\epsilon}{3} \, , & \qquad Y(e_{aR}) = -2 + \epsilon \, .
\end{array}
\label{hyper2}
\end{eqnarray}
It has been also noted that the nonstandard hypercharge piece is
proportional to $B-L$.  So right-handed neutrinos lead to
hypercharge dequantization via the nonstandard formula
\begin{equation}
Y = Y_{\rm SM} - \epsilon (B-L) \, , \label{dequantization}
\end{equation}
and then electric charge dequantization according to
\begin{equation}
Q = Q_{\rm SM} - \frac{\epsilon}{2} \; (B-L) \, .
\label{Qdequantization}
\end{equation}
The gauging of this combination does not affect the consistency of
the SM and its agreement with experiment is maintained if
$\epsilon$ is taken small enough.  Nevertheless, it has been
observed that an extension of the SM with Majorana masses to the
right-handed neutrinos restores charge quantization because
$\epsilon=0$, and $B-L$ is no more an exact symmetry.

We will demonstrate, however, that the realization of presymmetry
in the SM with Dirac neutrinos, the principle of weak
topological-charge confinement, and the weak topological character
of $B-L$, ensure the charge quantization and ungauging of $B-L$.

First we note that, due to the relation in Eq.~({\ref{BL}), charge
dequantization is via the topological $Z_3$ charge in both
quark--prequark and lepton--prelepton systems.  Secondly we remark
that this dequantization does not break presymmetry. In fact, the
dequantization of topological charge becomes evident once
Eq.~(\ref{newhyper}) is changed to include the $\epsilon$
parameter:
\begin{eqnarray}
\begin{array}{rcl}
Y(q) &=& \displaystyle Y(\ell) - \frac{4}{3} \; (1-\epsilon)
(3B-L)(\ell) \, , \\[12pt] Y(\ell) &=& \displaystyle Y(q) - \frac{4}{3}
\; (1-\epsilon) (3B-L)(q) \, ,
\end{array}
\label{hyper3}
\end{eqnarray}
with $Y$ as in Eq.~(\ref{dequantization}). Presymmetry still holds
and Eq.~(\ref{hathyper}) becomes modified in the same way:
\begin{eqnarray}
\begin{array}{rcl}
Y(q) &=& \displaystyle Y(\hat{q}) - \frac{4}{3} \; (1-\epsilon)
(B-3L)(\hat{q}) \, , \\[12pt] Y(\ell) &=& \displaystyle Y(\hat{\ell}) -
\frac{4}{3} \; (1-\epsilon) (B-3L)(\hat{\ell}) \, .
\end{array}
\label{hyper4}
\end{eqnarray}
On the other hand, Eq.~(\ref{hatanother}) is also altered as
follows:
\begin{eqnarray}
\begin{array}{rcl}
Y(q) - (1-\epsilon) (B-L)(q) &=& Y(\hat{q}) - (1-\epsilon)
(B-L)(\hat{q}) \, , \\[12pt] Y(\ell) - (1-\epsilon) (B-L)(\ell)
&=& Y(\hat{\ell}) - (1-\epsilon) (B-L)(\hat{\ell}) \, .
\end{array}
\label{hyper5}
\end{eqnarray}
From Eqs.~(\ref{deltaYl}), (\ref{deltaYq}), (\ref{hyper4}) and
(\ref{hyper5}), it is now clear that the dequantization of the
topological $Z_{3}$ charge takes the form
\begin{equation}
Q^{(3)} = (1-\epsilon) Q^{(3)}_{\rm SM} \, ,
\end{equation}
where $Q^{(3)}_{\rm SM}$ is defined in Eq.~(\ref{BL}) for
(pre)quarks and (pre)leptons and in Eq.~(\ref{Q3charge}) for the
four-instanton field configuration attached to prequarks and
preleptons.  However, this result contradicts our hypothesis that
$Q^{(3)}$ is a $Z_{3}$ number since $\epsilon$ is an arbitrary
small parameter. Hence $\epsilon=0$, which leads to charge
quantization and ungauging of $B-L$, with quarks and leptons
having the proper hypercharge and electric charge values. From
Eqs. (\ref{hyper3}) and (\ref{hyper4}) we also see that the value
of the topological index $n=4$, instead of $n=4(1-\epsilon)$,
physically means electric charge quantization.

\section{Fractional Charge Confinement}
\label{confinement}

Many strong arguments and techniques have been developed that have
elevated gauge confinement into an accepted principle.\cite{book2}
According to this principle, the only observable gauge states of
an exact Yang--Mills gauge theory with the property of asymptotic
freedom are those that are singlets under the corresponding gauge
symmetry.  This principle rests on the local action of non-Abelian
gauge fields.  There are no doubts that quark confinement inside
hadrons arises as a result of color confinement.  However, as
discussed in Sec.~\ref{Qquantization}, the charge allocation is
arbitrary if the SM is modified only by the addition of Dirac
right-handed neutrinos. It is not understood, for example, why
there are no fractionally charged hadrons made of colored quarks.
Clearly, the answer must be provided by the same theoretical
framework used to understand the origin of charge quantization.

Within our context, the no existence of free matter with
fractional electric charge is a logical consequence of the
principles of presymmetry and weak topological-charge confinement,
which explain charge quantization.  Starting from these principles
we have shown a dequantization of the quark electric charge
according to
\begin{equation}
\displaystyle Q(q) = Q(\hat{q}) + \frac{2}{3} \, Q^{(3)}(\hat{q})
\, ,
\end{equation}
where $Q^{(3)}(\hat{q})$ is defined in Eq.~(\ref{Q3charge}), then
providing a topological explanation of its fractional value. It
has a local part equal, by presymmetry, to the {\it integer} local
charge of the lepton partner. The other flavor-independent {\it
fractional} piece has a topological character which is induced by
an integer topological charge associated with a weak
four-instanton event and a modulo charge $Z_{3}$ coming from the
principle of weak topological-charge confinement. Both the
one-third and the order of the modulo charge of this topological
part are due to the number of quark colors. The no existence of
fractionally charged hadrons is therefore related to the absence
of topological effects on their physical properties. But, this
latter feature is guaranteed from the beginning by the principle
of weak topological-charge confinement.

In the quark model, hadrons are either bound states of one quark
and one antiquark (mesons) or states of three quarks (baryons). In
both of these cases we are left with the neutral $Q^{(3)}$ charge
and topologically trivial integer-charged particles:
\begin{eqnarray}
\begin{array}{rcl}
\displaystyle Q(q\bar{q}) &=& \displaystyle
Q(\hat{q}\hat{\bar{q}}) + \frac{2}{3} \,
Q^{(3)}(\hat{q}\hat{\bar{q}}) =
Q(\hat{q}\hat{\bar{q}}) \, , \\[12pt]
\displaystyle Q(qqq) &=& \displaystyle Q(\hat{q}\hat{q}\hat{q}) +
\frac{2}{3} \, Q^{(3)}(\hat{q}\hat{q}\hat{q}) =
Q(\hat{q}\hat{q}\hat{q}) + 2 \, .
\end{array}
\end{eqnarray}
In a sense the electric charge of hadrons is due to the integer
electric charge of prequarks and in the case of baryons with two
extra units of universal character.  No fractionally charged
hadrons can be produced. It is worth remarking, however, that the
topological-charge confinement is secondary to the gauge
confinement as electroweak forces by themselves cannot lead to
actual confinement. Quarks are the physical particles and their
confinement is due to their strong color interactions, but
presymmetry and the secondary weak topological-charge confinement
account for their fractional charges, so explaining why there are
no fractionally charged hadrons.  We may say that fractional
electric charge is also confined.

\section{Conclusions}
\label{conclusions}

We have answered fundamental questions unresolved by the SM with
Dirac right-handed neutrinos relying on a hidden discrete
electroweak symmetry between quarks and leptons, a $Z_2$ symmetry
which has been formalized with a principle of electroweak
quark--lepton symmetry; there is no such a symmetry in the SM
because neutrinos are taken massless and no right-handed neutrinos
are introduced. This symmetry is inferred from the one-to-one
constraints given in Eq.~(\ref{hyper}) which, if not accidental,
exhibit integer (fractional) hypercharges underlying quarks
(leptons) and a global $4/3$ shift, independent of flavor and
handness, associated with baryon and lepton numbers as in
Eq.~(\ref{newhyper}). To understand this pattern, the new quark
(lepton) states of prequarks (preleptons) with integer
(fractional) charges have been added to the SM with Dirac
neutrinos. The discrete prequark--lepton (quark--prelepton)
presymmetry generates the hidden $Z_2$ symmetry. A principle of
weak topological-charge confinement, which is secondary to the
strong gauge confinement of colored quarks, has also been stated,
guaranteeing that weak topological effects do not show up at the
level of observable particles at zero temperature. And using
well-known topological properties of gauge fields, we have found
that topological quarks and topological preleptons have underlying
integer topological charges, associated with an additive group
$Z_3$, whose modulo order is related to the number of quark colors
and quark--lepton families. These topological charges induce the
exposed charge shift and the breaking of presymmetry. Our
conclusion is that the primordial electroweak gauge-field
difference between topological quarks and leptons is the
nonequivalence between the topological vacua of their weak field
configurations, produced by a weak four-instanton which carries
the topological charge, induces the universal fractional piece of
charge needed for charge normalization, and breaks presymmetry.

To make clear the true character of the elements of our model, we
note that prequarks and preleptons, and so the charge structure of
confined quarks, have no direct experimental manifestations;
leptons have no charge structure. Neither prequarks and preleptons
are new particles with definite mass values nor the associated
presymmetry has a mass scale breaking. Moreover, preleptons are
neither observable nor confined particles. Prequarks and
preleptons are simply massless timelike pre-states of quarks and
leptons, respectively, which may be seen as suitable mathematical
entities out of which to build up the actual particle states. The
model of prequarks and preleptons is not to be taken literally as
a dynamical model such that their interactions can determine
properties of quarks and leptons as mass and mixing angles.
Although our starting point at the bare level considers prequarks
and preleptons as interacting with the standard gauge and Higgs
fields in the usual way, with a Lagrangian that looks the same as
that of the SM with quarks and leptons excepting hypercharge
couplings and incorporation of Dirac right-handed neutrinos, they
are not the particles that do the job with the physical $W^{\pm}$
and $Z^{0}$ weak bosons, photons, gluons, and Higgs boson of the
SM. Prequarks and preleptons are no physical states and even no
gauge proper states due to gauge quantum anomalies. It is worth
recalling that in the original model of quarks, they were
presented as mathematical unities. They ended up, however, as real
dynamical particles. We now claim that prequarks and preleptons
are indeed the authentic mathematical entities underlying the
physical particle states.

Nevertheless, our model has several phenomenological implications.
It explains the fractional charge of quarks and the quark--lepton
charge relations. It predicts that the number of fermion
generations has to be equal to the number of quark colors. It
predicts $B-L$ conservation and so the ungauging of $B-L$ and the
Dirac character of massive neutrinos. It accounts for the
topological charge conservation in quantum flavor dynamics.
Finally, it explains charge quantization and the no observation of
fractionally charged hadrons.

We have argued that the charge normalization to SM values,
demanded by anomaly cancellation, and the zero topological charge
in leptons and hadrons effectively remove the extremely large time
scale associated with weak topological effects, occurring in
general if the weak gauge fields are assumed to remain in the
small coupling regime, i.e. in the phase of the zero-temperature
scale of the model.

Our results certainly mean an improvement of the conceptual
foundations of the SM; if presymmetry is not recognized as a
hidden symmetry in the SM modified by Dirac right-handed
neutrinos, then the number of generations cannot be fixed, charge
allocation can be arbitrary though restricted by the anomaly
cancellation and nonvanishing of fermion masses, $B-L$ can be
gauged, and the absence of fractionally charged hadrons cannot be
understood, despite the quark confinement due to QCD. However, a
new physics would remain to be discovered to understand any
dynamics that may address many other problems left open, such as
the mass hierarchies.

On the other hand, the model may need extensions to be further
substantiated and have indicated indirectly new experimentally
observable predictions. The main motivation comes from the fact
that the hidden presymmetry relates quarks with leptons which do
not have the strong QCD interactions, so that the $Z_2$ symmetry
is not a property of the whole Lagrangian of the system. A
particularly promising idea that oversteps this limitation and
leads to a rich phenomenology beyond the SM is to extend
presymmetry from matter to forces, so that symmetric bare quarks
and leptons interact with symmetric gauge bosons, i.e. follow the
principle of electroweak quark--lepton symmetry in its strongest
form. This would require symmetric exotic quarks and leptons as
those assumed in Ref.~\refcite{Matute4}, which allow to have a
$Z_2$ symmetry over the strong interaction sector and a symmetric
duplication of the gauge group of the SM and its unifying
expansions. It is worth noting here that the three new families of
quarks must have a new color degree of freedom in order to not
contradict the conclusion that the number of generations and the
number of quark colors have to be equal, although the
corresponding exotic matter must decay into ordinary matter
rapidly enough to not contravene the cosmological evidences on the
absence of exotic baryons. We are currently considering this
interesting presymmetric extension and hope to report on our
findings elsewhere.  In addition, our study may be further
extended to the high-temperature phase of the model, at the scale
of the electroweak transition, where the weak topological effects
become important without violating the principle of
topological-charge confinement.

\section*{Acknowledgments}

We thank J. Gamboa, M. S. Plyushchay and L. Vergara for carefully
reading the manuscript and providing useful comments. This work
was supported in part by the Departamento de Investigaciones
Cient\'{\i}ficas y Tecnol\'ogicas, Universidad de Santiago de
Chile, USACH.

\end{document}